\documentclass[onecolumn,showpacs,%
  nofootinbib,aps,superscriptaddress,%
  prd,notitlepage,showkeys,10pt]{revtex4-1}

\usepackage{amssymb}
\usepackage{amsmath}
\usepackage{graphicx}
\usepackage{dcolumn}
\usepackage{hyperref}
\usepackage{lineno}
\begin{document}

\title{Protocol for counterfactually transporting an unknown qubit}
\author{Hatim Salih}
\email[]{salih.hatim@gmail.com}
\affiliation{Qubet Research, London NW6 1RE, UK}

\date{\today}

\begin{abstract}
%\linenumbers
Quantum teleportation circumvents the uncertainty principle using dual channels: a quantum one consisting of previously-shared entanglement, and a classical one, together allowing the disembodied transport of an unknown quantum state over distance. It has recently been shown that a classical bit can be counterfactually communicated between two parties in empty space, ``Alice'' and ``Bob''. Here, by using our ``dual'' version of the chained quantum Zeno effect to achieve a counterfactual CNOT gate, we propose the first protocol for transporting an unknown qubit counterfactually, that is without any physical particles travelling between Alice and Bob---no classical channel and no previously-shared entanglement.

\end{abstract}


\maketitle
\section{Introduction}

% For Original Research Articles, Clinical Trial Articles, and Technology Reports the introduction should be succinct, with no subheadings.
%
% For Clinical Case Studies the Introduction should include symptoms at presentation, physical exams and lab results.
%
In contrast to classical information, quantum states cannot be faithfully copied---as proven by the no-cloning theorem \cite{Wooters}. In fact doing so would not only violate the uncertainty principle, but would also violate the rule against faster-than-light signalling \cite{Gisin}. Teleporting an unknown qubit, whereby an identical version appears elsewhere, was presumed to firmly belong to the realm of science fiction---until Bennett et al. \cite{Bennett} showed it possible using previously shared entanglement and a classical channel, such as a phone line. The original qubit, in accordance with no-cloning, is duly destroyed in the process. Quantum teleportation has been extended to systems with continuous variables \cite{Vaidman 1994}, and has been demonstrated in numerous key experiments \cite{Zeilinger, Braunstein, Riebe, Barrett, Sherson}. 

Here, we wonder: Is the disembodied transport of an unknown qubit over distance possible, even in principle, without recourse to previously-shared entanglement or a classical channel?---and intriguingly, without physical particles travelling between Alice and Bob? The answer, as we show, is  surprisingly yes.

It has recently been shown that a classical bit can be counterfactually transferred between two parties in empty space \cite{Salih, ResearchHighlight}. The key ideas behind direct counterfactual quantum communication---a version of which has recently been realised experimentally \cite{Cao}---are interaction-free measurement \cite{Dicke, Elitzur} and the quantum Zeno effect \cite{Kwiat 1995, Kwiat 1999, Misra, Asher, Agarwal}. Interaction-free measurement is based on the fact that the presence of a blocking object inside an interferometer destroys interference even if no particle hits the object. This has the interesting consequence that sometimes the presence of such an object can be inferred without the object directly interacting with any particles. Noh \cite{Noh} used this to design a counterfactual quantum key distribution protocol whereby, for shared random bits, no information-carrying photons travel between Alice and Bob. The quantum Zeno effect on the other hand is based on the fact that repeated measurement of an evolving quantum system inhibits its evolution, ``freezing'' it in its initial state---an effect reminiscent of the proverbial watched kettle that never boils. The quantum Zeno effect can arbitrarily boost the efficiency of interaction-free measurements.

%\begin{methods}
\section{Methods}

\begin{figure}
\centering
\includegraphics[width=0.5\textwidth]{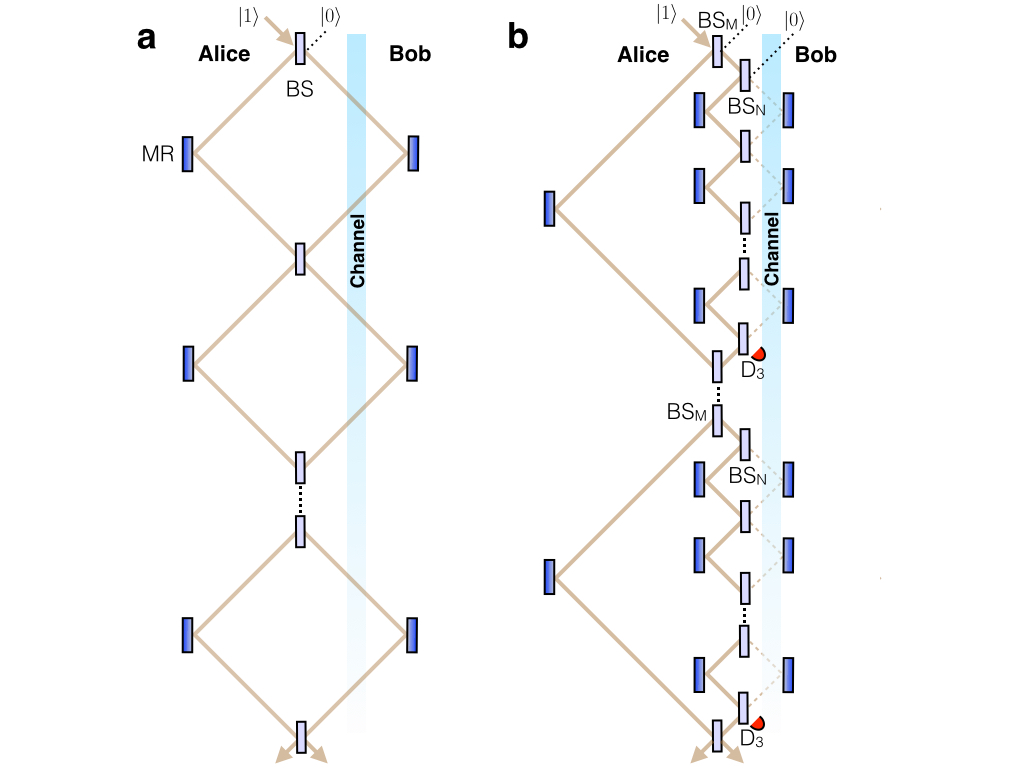}
\caption{\label{fig: MZ-CQZE}{\bf Proposal for counterfactual Mach-Zehnder CNOT.} Bob effects a quantum superposition of blocking and not blocking the channel, not practical for this Mach-Zehnder scenario, which acts as the CNOT’s control qubit. {\bf a}, Partially counterfactual CNOT gate. Beamsplitters $BS$ are highly reflective, with reflectivity $R=\cos^2{\pi /2N}$, where $N$ is the total number of $BS$s. The gate, however, only works for one of Alice's inputs (``0''). Moreover, it is not counterfactual for the part of the superposition where Bob does not block, in which case the photon passes through the channel. {\bf b}, Fully counterfactual CNOT gate based on the chained quantum Zeno effect (CQZE). Between successive $BS_M$s, of which there are $M$, there are $N$ beamsplitters $BS_N$. While the scheme only works for Alice's ``0'' input, complete counterfactuality is ensured as any photon going into the channel would be lost due to measurement by Bob's object or else end up at one of the detectors $D_3$: the chained quantum Zeno effect. For large enough $M$ and $N$, the probability amplitude of the photon being in the channel is virtually zero.}
\end{figure}

We start by generalising the action of the chained quantum Zeno effect (CQZE) \cite{Hosten} of the Mach-Zehnder form in reference \cite{Salih} to the case of Bob effecting a quantum superposition of blocking and not blocking the channel, showing first how it can function as a fully counterfactual, though not yet practical, CNOT gate, for only one of Alice's input states. Consider the Mach-Zehnder interferometry Zeno setup of FIG. \ref{fig: MZ-CQZE}a. The first concept we require here is Bob effecting a quantum superposition of blocking and not blocking the transmission channel \cite{Dicke, Elitzur, Kwiat 1999}. Although this is easier to imagine from a practical point of view for the Michelson version we discuss later in this section, this Mach-Zehnder version is easier to explain. Here, $BS$ stands for beamsplitter. The action of $BS$ on Alice's photon is the following, $\left| \text{10} \right\rangle \to \cos \theta \left| \text{10} \right\rangle +\sin \theta \left| \text{01} \right\rangle$ and $\left| \text{01} \right\rangle \to \cos \theta \left| \text{01} \right\rangle -\sin \theta \left| \text{10} \right\rangle$, where the state $\left| \text{10} \right\rangle$ corresponds to the photon being on the left of $BS$, the state $\left| \text{01} \right\rangle$ corresponds to the photon being on the right of $BS$, and $\cos \theta =\sqrt{R}$, with $R$ being the reflectivity of $BS$. We set $\theta =\pi /2N$, where $N$ is the number of beamsplitters. Let the initial combined state of Bob's quantum object together with Alice's photon, impinging on the first beamsplitter $BS$ from the top left, be $(\alpha \left| \text{pass} \right\rangle + \beta \left| \text{block} \right\rangle) \otimes \left| \text{10} \right\rangle$. The state $\left| \text{pass} \right\rangle$ corresponds to Bob's object not blocking the channel, and has probability amplitude $\alpha$. The state $\left| \text{block} \right\rangle$ corresponds to Bob's object blocking the channel, and has probability amplitude $\beta$. 

In the following, starting from the second $BS$ we post-select before each $BS$ the state corresponding to the photon not hitting Bob's object. If the photon hits Bob's object we assume it is absorbed. But the exact nature of the interaction does not matter, only that the photon is lost; the protocol fails. Immediately after the first $BS$ the combined state is rotated to $(\alpha \left| \text{pass} \right\rangle + \beta \left| \text{block} \right\rangle) \otimes (\cos \theta \left| \text{10} \right\rangle + \sin \theta \left| \text{01} \right\rangle)$, which immediately before the second $BS$, given the photon is not lost to Bob's object, becomes $\alpha \left| \text{pass} \right\rangle \otimes (\cos {\theta} \left| \text{10} \right\rangle +\sin {\theta} \left| \text{01} \right\rangle) + \beta \left| \text{block} \right\rangle \otimes \cos \theta \left| \text{10} \right\rangle$. And immediately after the second $BS$ this state is rotated to $\alpha \left| \text{pass} \right\rangle \otimes (\cos 2{\theta}\left| \text{10} \right\rangle +\sin 2{\theta} \left| \text{01} \right\rangle) + \beta \left| \text{block} \right\rangle \otimes {\cos} \theta(\cos \theta \left| \text{10} \right\rangle +\sin {\theta} \left| \text{01} \right\rangle)$, which immediately before the third $BS$, given the photon is not lost to Bob's object, becomes $\alpha \left| \text{pass} \right\rangle \otimes (\cos 2{\theta}\left| \text{10} \right\rangle +\sin 2{\theta} \left| \text{01} \right\rangle) + \beta \left| \text{block} \right\rangle \otimes {\cos}^2 \theta \left| \text{10} \right\rangle$. And immediately after the third $BS$, this state is rotated to $\alpha \left| \text{pass} \right\rangle \otimes (\cos 3{\theta}\left| \text{10} \right\rangle +\sin 3{\theta} \left| \text{01} \right\rangle) + \beta \left| \text{block} \right\rangle \otimes {\cos}^2 \theta(\cos \theta \left| \text{10} \right\rangle +\sin {\theta} \left| \text{01} \right\rangle)$.

Thus after $n$ beamsplitters the combined state has evolved as,

\begin{equation}
\begin{split}
&(\alpha \left| \text{pass} \right\rangle + \beta \left| \text{block} \right\rangle) \otimes \left| \text{10} \right\rangle \to
\\ &\alpha \left| \text{pass} \right\rangle \otimes (\cos n{\theta}\left| \text{10} \right\rangle +\sin n{\theta} \left| \text{01} \right\rangle) + 
\\ &\beta \left| \text{block} \right\rangle \otimes {\cos}^{n-1} \theta(\cos \theta \left| \text{10} \right\rangle +\sin {\theta} \left| \text{01} \right\rangle).
\end{split}
\end{equation}

And after $N$ beamsplitters, with $N$ very large, the combined state of Bob's quantum object and Alice's photon becomes $\alpha \left| \text{pass} \right\rangle \left| \text{01} \right\rangle + \beta \left| \text{block} \right\rangle \left| \text{10} \right\rangle$. We thus have a CNOT gate with Bob's as the control bit, $\left| \text{block} \right\rangle \equiv \left| \text{0} \right\rangle$, $\left| \text{pass} \right\rangle \equiv \left| \text{1} \right\rangle$, and Alice's as the target bit, $\left| \text{10} \right\rangle \equiv \left| \text{0} \right\rangle$, $\left| \text{01} \right\rangle \equiv \left| \text{1} \right\rangle$, albeit for only one of Alice's possible input states, namely $\left| \text{0} \right\rangle$. Moreover, the scheme is only counterfactual for the part of the superposition corresponding to Bob blocking and is not counterfactual for the part of the superposition corresponding to Bob not blocking, where Alice's photon gradually ``leaks'' into the channel.  

We now show how to achieve complete CNOT counterfactuality, for Alice's input state $\left| \text{0} \right\rangle$, using the chained quantum Zeno effect (CQZE) setup of FIG. \ref{fig: MZ-CQZE}b. Here, Alice's photon goes through $M$ beamsplitters $B{{S}_{M}}$, with ${{\theta }_{M}}=\pi /2M$. Between successive $BS_M$s the photon goes through $N$ beamsplitters $BS_N$, with ${{\theta }_{N}}=\pi /2N$. The state $\left| \text{100} \right\rangle$ corresponds to Alice's photon being on the left of $BS_M$, the state $\left| \text{010} \right\rangle$ corresponds to the photon being on the right of $BS_M$ and on the left of $BS_N$, and the state $\left| \text{001} \right\rangle$ corresponds to the photon being on the right of $BS_N$. For the $m$-th cycle,

\begin{equation}
\begin{split}
\label{For m-th}
&(\alpha \left| \text{pass} \right\rangle + \beta \left| \text{block} \right\rangle) \otimes \left| \text{010} \right\rangle \to 
\\ &\alpha \left| \text{pass} \right\rangle \otimes (\cos n{{\theta}_{N}}\left| \text{010} \right\rangle +\sin n{{\theta}_{N}} \left| \text{001} \right\rangle) + 
\\ &\beta \left| \text{block} \right\rangle \otimes {\cos}^{n-1}{\theta}_{N} (\cos {{\theta}_{N}} \left| \text{010} \right\rangle +\sin {{\theta}_{N}} \left| \text{001} \right\rangle).
\end{split}
\end{equation}

And after $N$ beamsplitters $BS_N$, with $N$ very large, the combined state of Bob's quantum object and Alice's photon $\simeq$ $\alpha \left| \text{pass} \right\rangle \left| \text{001} \right\rangle + \beta \left| \text{block} \right\rangle \left| \text{010} \right\rangle$. But Alice's single photon is initially in the state $\left| \text{100} \right\rangle$, as shown in FIG. \ref{fig: MZ-CQZE}b, with all unused ports in the vacuum state. After the $m$-th $BS_M$,

\begin{equation}
\begin{split}
\label{After m-th}
&(\alpha \left| \text{pass} \right\rangle + \beta \left| \text{block} \right\rangle) \otimes \left| \text{100} \right\rangle \to 
\\ &\alpha \left| \text{pass} \right\rangle \otimes {\cos}^{m-1}{\theta}_{M} (\cos {{\theta}_{M}}\left| \text{100} \right\rangle +\sin {{\theta}_{M}} \left| \text{010} \right\rangle) + 
\\ &\beta \left| \text{block} \right\rangle \otimes (\cos m{{\theta}_{M}} \left| \text{100} \right\rangle +\sin m{{\theta}_{M}} \left| \text{010} \right\rangle).
\end{split}
\end{equation}

And after the $M$-th $BS_M$, with $M$ very large, the combined state of Bob's quantum object and Alice's photon $\simeq$ $\alpha \left| \text{pass} \right\rangle \left| \text{100} \right\rangle + \beta \left| \text{block} \right\rangle \left| \text{010} \right\rangle$. We thus have a fully counterfactual CNOT gate, with Bob's as the control bit, $\left| \text{pass} \right\rangle \equiv \left| \text{0} \right\rangle$, $\left| \text{block} \right\rangle \equiv \left| \text{1} \right\rangle$, and Alice's as the target bit, $\left| \text{100} \right\rangle \equiv \left| \text{0} \right\rangle$, $\left| \text{010} \right\rangle \equiv \left| \text{1} \right\rangle$, again for only one of Alice's possible input states, namely $\left| \text{0} \right\rangle$. 

Complete counterfactuality is ensured: Any photon going into the channel would either be lost due to measurement by Bob's object or else end up at one of the detectors $D_3$. This is the most direct, intuitive definition of counterfactuality. Defining counterfactuality in terms of weak measurements, for instance, has proven controversial \cite{Vaidman 2014, Salih reply}, leading to the claim that ``the photon did not enter the channel, it did not leave, but it was there''. (Weak measurements are also known to record faster-than-light speeds.) Moreover, by Eqs. \ref{For m-th} and \ref{After m-th}, the probability amplitude of the photonic state $\left| \text{001} \right\rangle$ corresponding to the photon being in the channel is virtually zero for large enough $M$ and $N$. Nevertheless, the scheme is not practical, and more fundamentally, it only works for one of Alice's input states.

\begin{figure}
\centering
\includegraphics[width=0.5\textwidth]{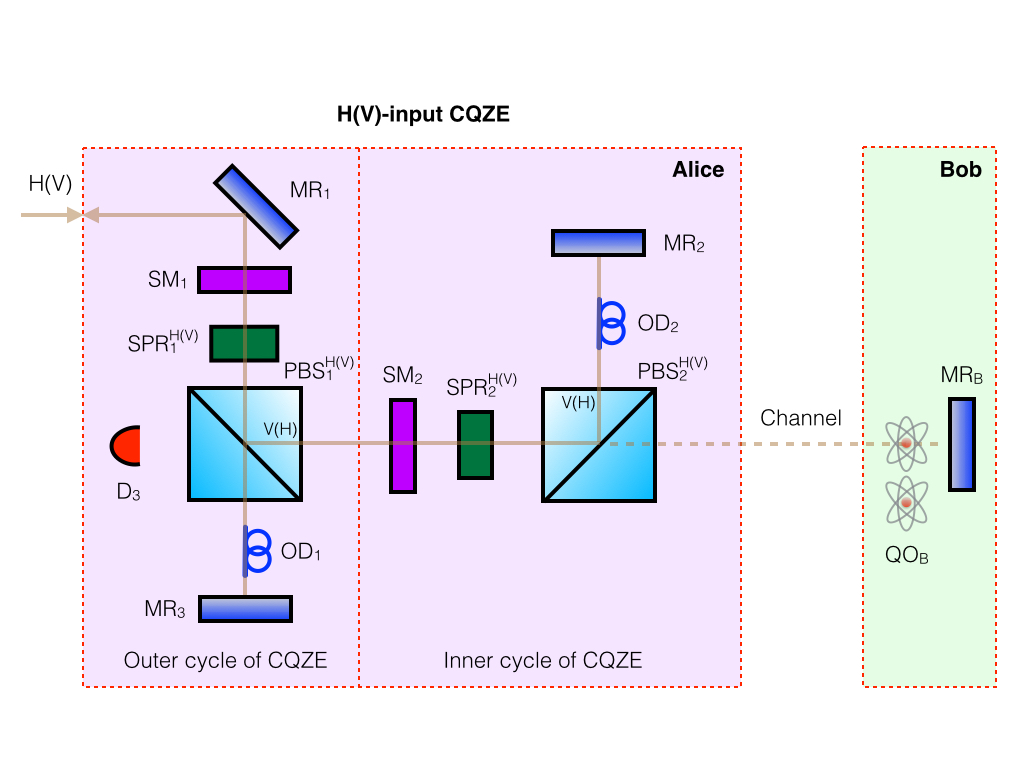}
\caption{\label{fig: M-CQZE}{\bf Proposal for fully counterfactual CNOT gate based on Michelson CQZE.} By using polarising beamsplitter $PBS^{H(V)}$ that passes $H(V)$ photons and reflects $V(H)$, this setup can take $H(V)$ input from Alice, i.e. ``0''(``1''), but not yet a superposition. Bob implements his qubit as a superposition of blocking and not blocking the channel using his quantum object $QO_B$. Switchable mirror $S{{M}_{1(2)}}$ is initially turned off allowing the photon in, but is then turned on for $M(N)$ outer(inner) cycles before it is turned off again, allowing the photon out. The combined action of switchable polarisation rotators $SPR$ and polarising beamsplitters $PBS$ achieves the function of beamsplitters $BS$ in the Mach-Zehnder version of FIG \ref{fig: MZ-CQZE}. $MR$ stands for mirror, and $OD$ for optical delay. Again, complete counterfactuality is ensured as any photon going into the channel would either be lost due to measurement by Bob's object $QO_B$ or else end up at detector $D_3$: the chained quantum Zeno effect. For large enough $M$ and $N$, the probability amplitude of the photon being in the channel is virtually zero.}
\end{figure}

Let us now consider a more versatile Michelson version, showing in principle how to achieve a counterfactual quantum CNOT gate for all possible input values. Here, the function of $BS$ in the Mach-Zehnder CQZE setup of FIG. 1b is achieved by the combined action of switchable polarisation rotator $SPR$ and polarising beamsplitter \cite{Salih, Salih PRA} $PBS$, as shown in FIG. \ref{fig: M-CQZE}. Bob's quantum object $QO{_B}$ exists in a superposition of blocking and not blocking the channel. Here, $H$($V$) refers to horizontal (vertical) polarisation. The action of $SPR^{H(V)}_i$, in Alice's $H(V)$-input setup, on her photon is the following, 

\begin{equation}
\left| H(V) \right\rangle \to \cos {{\theta}_i}\left| H(V) \right\rangle +\sin{{\theta}_i}\left| V(H) \right\rangle.
\end{equation}

\begin{equation}
\left| V(H) \right\rangle \to \cos {{\theta}_i}\left| V(H) \right\rangle -\sin {{\theta}_i}\left| H(V) \right\rangle.
\end{equation}

with $i=1, 2$ corresponding to $SPR$s with different rotation angles. We set rotation angle ${\theta}_{1(2)}=\pi/2M(N)$, with $SPR_{1(2)}$ switched on once per cycle when the photon, or part of it, is moving in the direction from $SM_{1(2)}$ towards $PBS_{1(2)}$. (This is in contrast to Salih et al. \cite{Salih} where switchable polarisation rotators are switched on when the photon is moving in the direction from switchable mirror $SM_{1(2)}$ towards $PBS_{1(2)}$ and again on the way back, leading to an undesired rotation in the last outer cycle, tiny for large $M$. The current scheme avoids this undesired rotation and resulting error altogether.) Switchable mirror $S{{M}_{1(2)}}$ is initially turned off allowing the photon in, but is then turned on for $M(N)$ cycles before it is turned off again, allowing the photon out.

The Michelson CQZE setup of FIG. \ref{fig: M-CQZE} takes $H(V)$ polarised photons as input, with $PBS^{H(V)}$ passing $H(V)$ photons and reflecting $V(H)$ as shown. Alice sends an $H(V)$ photon into the $H(V)$-input CQZE setup. By a similar evolution to Eqs. 2 and 3, for the part of the superposition corresponding to Bob not blocking the channel Alice's exiting photon is $H(V)$ polarised, while for the part of the superposition corresponding to Bob blocking the channel Alice's exiting photon is $V(H)$ polarised.

This means that Alice can encode her bit using polarisation. She encodes a ``0''(``1'') by sending an $H(V)$ photon into the corresponding $H(V)$-input CQZE setup. But can Alice encode a quantum superposition of ``0'' and ``1''? 

\begin{figure}
\centering
\includegraphics[width=0.5\textwidth]{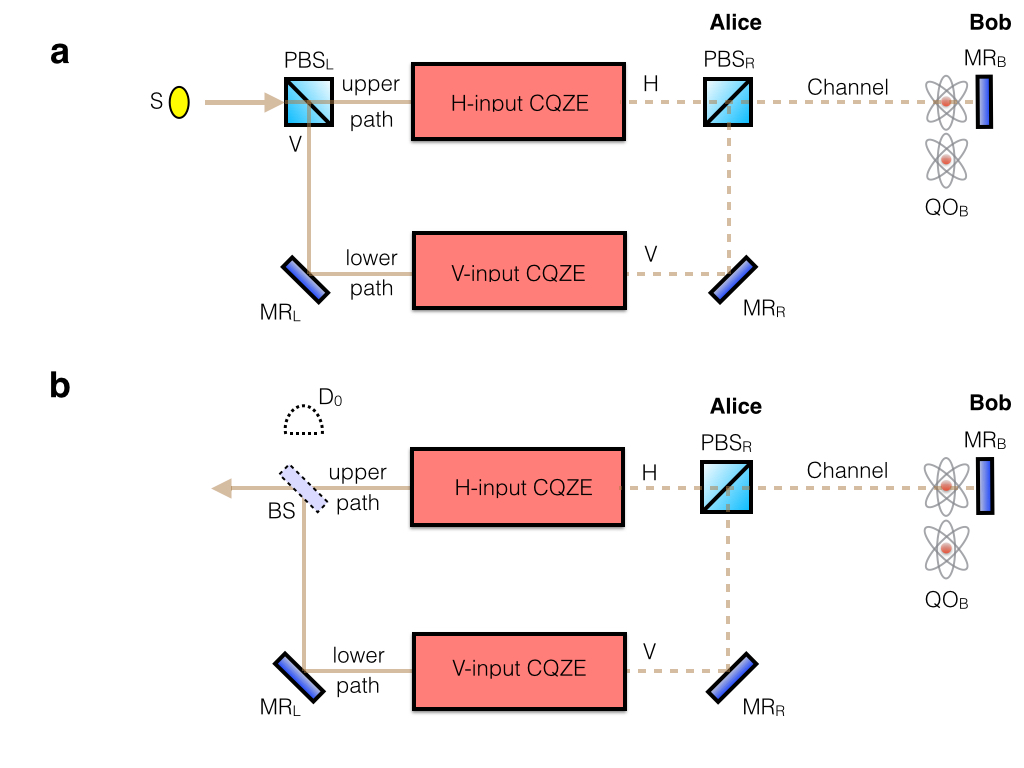}
\caption{\label{fig: Dual-CQZE}{\bf Proposal for fully counterfactual general input CNOT gate based on a dual CQZE.} {\bf a}, Alice sends a photon in the qubit state $\alpha \left| \text{H} \right\rangle + \beta \left| \text{V} \right\rangle$, while Bob's quantum object $QO_B$ is in the qubit state $\lambda \left| \text{pass} \right\rangle + \mu \left| \text{block} \right\rangle$. Alice's incoming photon is first separated into $H$ and $V$ components using polarising beamsplitter $PBS_L$, which are then respectively fed into the $H$-input and $V$-input CQZE modules from FIG. \ref{fig: M-CQZE}. {\bf b}, For Alice's exiting photon, $PBS_L$ is replaced by a 50:50 beamsplitter $BS$. If it is not detected at $D_0$, then the photon exits towards the left in the correct state. For the case of Alice initially sending an $H$ photon, as in the first step of our protocol, there is no need for BS, the photon exits towards the left in the correct state. $S$ stands for single-photon source.}
\end{figure}

Crucially, the answer is yes. She first passes her photon through $PBS_L$ in order to separate it into $H$ and $V$ components as shown in FIG. \ref{fig: Dual-CQZE}a. The $H$($V$) component is then fed into the corresponding $H(V)$-input CQZE setup. Bob can block or not block the transmission channel---or a quantum superposition of blocking and not blocking---for both $H$ and $V$ components which are recombined using $PBS_R$, FIG. \ref{fig: Dual-CQZE}a. The polarisation of Alice's exiting photon is determined by Bob's bit choice. This is our dual chained quantum Zeno effect.

Starting with Eqs. \ref{For m-th} and \ref{After m-th}, and adapting for the Michelson CQZE setup of FIG. \ref{fig: M-CQZE}, a lower bound of the probability that the photon avoids detection by detector $D_3$ in all cycles is $({1-{\left | \alpha \right |}^2{\sin}^2{\theta}_{M}})^{{M}}$. A lower bound of the probability that the photon avoids being lost due to measurement by Bob's object in all cycles is $\prod_{m=1}^{M} (1-{\left | \beta \right |}^2{\sin}^{2}{m\theta}_{M}{\sin}^{2}{\theta}_{N})^N$. Thus the ideal case efficiency of this counterfactual CNOT gate is at least,

\begin{equation}
\begin{split}
\label{Max efficiency}
({1-{\left | \alpha \right |}^2{\sin}^2{\theta}_{M}})^{{M}} \prod_{m=1}^{M} (1-{\left | \beta \right |}^2{\sin}^{2}{m\theta}_{M}{\sin}^{2}{\theta}_{N})^N.
%&\lambda(\alpha \left| \text{pass} \right\rangle \left| \text{H} \right\rangle + \beta \left| \text{block} \right\rangle \left| \text{V} \right\rangle) \otimes \left| \text{upper path} \right\rangle + \\
%&\mu(\alpha \left| \text{pass} \right\rangle \left| \text{V} \right\rangle + \beta \left| \text{block} \right\rangle \left| \text{H} \right\rangle) \otimes \left| \text{lower path} \right\rangle.
\end{split}
\end{equation} 
 
\begin{figure}
\centering
\includegraphics[width=0.5\textwidth]{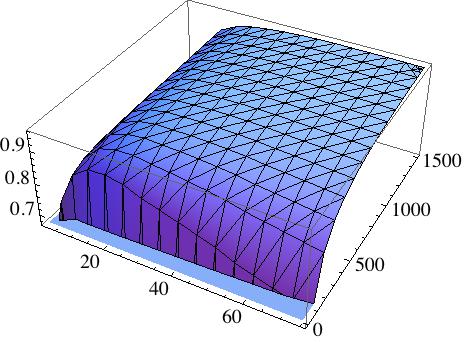}
\caption{\label{fig: Mesh}{\bf Ideal case efficiency of counterfactual CNOT gate.} The efficiency of an ideally implemented counterfactual CNOT gate, with Alice encoding a quantum superposition of ``0'' and ``1'', plotted against the number of outer and inner cycles M and N, with M up to 75, and N up to 1500. Bob's control qubit is assumed to be an equal superposition here. Efficiency approaches unity for $N \gg M \gg 1$. Errors or loss caused by imperfect optical elements and by noise in the channel, as well as decoherence of Bob's quantum object, all of which have an adverse effect, are ignored here.}
\end{figure}

In FIG. \ref{fig: Mesh} we plot this ideal case efficiency for different $M$s and $N$s, for $\alpha = \beta = 1/\sqrt2$. Given perfect implementation, we see that efficiency approaches unity for $N \gg M \gg 1$. For instance for $M=50$ and $N=1250$, ideal efficiency is already $95\%$. 

Complete counterfactuality is ensured: Any photon going into the channel would either be lost due to measurement by Bob's object or else end up at detector D3. Moreover, the probability amplitude of the photonic state corresponding to the photon being in the channel is virtually zero for large enough $M$ and $N$.

%Unlike the Mach-Zehnder scenario discussed above where, for Bob not blocking, the last $BS_M$ causes an undesired rotation---tiny for large $M$---in this Michelson implementation, a small number of outer cycles $M$ does not lead to output errors from the counterfactual CNOT gate, but would instead lead to more instances of the gate failing through photon loss. However, a small number of inner cycles $N$ would lead to output errors as well as more photon loss, as would imperfect implementation.

Reference \cite{Salih}, FIG. 4, contains an analysis for the case of transferring classical bits counterfactually in the presence of imperfections, for the two cases of Bob blocking the channel and that of Bob not blocking the channel. We see that the CQZE is sensitive to rotation errors by polarisation rotators $SPR$ as well as noise randomly blocking the channel. We expect a similar effect for our counterfactual quantum CNOT gate. A detailed analysis in the presence of imperfections, more crucial for a large number of cycles implementation, is left to a future study.

Note that in their experimental implementation of the CQZE of reference \cite{Salih}, in the single-photon regime, Cao et al. \cite{Cao} managed to mimic the action of switchable mirrors, whose direct realisation would have otherwise been challenging, by using slightly tilted semi-reflective mirrors. They post-selected only the photons that emerged in the correct spatial and temporal modes corresponding to the desired number of cycles.     

Now, consider the most general case where Alice sends a photon in the superposition $\lambda \left| \text{H} \right\rangle + \mu \left| \text{V} \right\rangle$, with Bob's object in the superposition $\alpha \left| \text{pass} \right\rangle + \beta \left| \text{block} \right\rangle$. We get the following superposition for Alice's exiting photon, from the upper path $H$-input CQZE module and the lower path $V$-input CQZE module, FIG. \ref{fig: Dual-CQZE}b,

\begin{equation}
\begin{split}
\label{Exiting superposition}
&\lambda(\alpha \left| \text{pass} \right\rangle \left| \text{H} \right\rangle + \beta \left| \text{block} \right\rangle \left| \text{V} \right\rangle) \otimes \left| \text{upper path} \right\rangle + \\
&\mu(\alpha \left| \text{pass} \right\rangle \left| \text{V} \right\rangle + \beta \left| \text{block} \right\rangle \left| \text{H} \right\rangle) \otimes \left| \text{lower path} \right\rangle.
\end{split}
\end{equation}
 
All we need now is to combine the two photonic states from the upper and lower paths. This is done by replacing $PBS_L$ in FIG. \ref{fig: Dual-CQZE}a by a 50:50 beamsplitter $BS$, as shown in FIG. \ref{fig: Dual-CQZE}b. We define the upper path as above or to the right of $BS$, and the lower path as below or to the left of $BS$. Let's rename the states $(\alpha \left| \text{pass} \right\rangle \left| \text{H} \right\rangle + \beta \left| \text{block} \right\rangle \left| \text{V} \right\rangle)$ and $(\alpha \left| \text{pass} \right\rangle \left| \text{V} \right\rangle + \beta \left| \text{block} \right\rangle \left| \text{H} \right\rangle)$ as $\left| \nwarrow \right\rangle$ and $\left| \swarrow \right\rangle$ respectively. We can rewrite the exiting state, Eq. \ref{Exiting superposition}, as $(\lambda \left| \nwarrow \right\rangle \left| \text{upper path} \right\rangle + \mu \left| \swarrow \right\rangle \left| \text{lower path} \right\rangle)$. Feeding this state into $BS$, which applies a $\pi/2$-rotation to the path qubit, gives, 

\begin{equation}
\begin{split}
\label{Exiting BS}
&1/\sqrt2(\lambda \left| \nwarrow \right\rangle + \mu \left| \swarrow \right\rangle) \otimes \left| \text{lower path} \right\rangle + \\
&1/\sqrt2(\lambda \left| \nwarrow \right\rangle - \mu \left| \swarrow \right\rangle) \otimes \left| \text{upper path} \right\rangle.
\end{split}
\end{equation}

which means we can obtain the desired state $\lambda \left| \nwarrow \right\rangle + \mu \left| \swarrow \right\rangle$ with $50\%$ probability upon measuring the path qubit. (We will shortly deal with the other $50\%$.) This measurement is carried out at $D_0$,  FIG. \ref{fig: Dual-CQZE}b, without destroying the photon when ideal \cite{Nogues, Haroche}. If the photon is not detected there we know it is in the other path travelling towards the left, in the correct state. So starting with the most general input states we have got,

\begin{equation}
\begin{split}
&(\alpha \left| \text{pass} \right\rangle + \beta \left| \text{block} \right\rangle) \otimes (\lambda \left| \text{H} \right\rangle + \mu \left| \text{V} \right\rangle) \to
\\ &\lambda (\alpha \left| \text{pass} \right\rangle \left| \text{H} \right\rangle + \beta \left| \text{block} \right\rangle \left| \text{V} \right\rangle) + \\
&\mu (\alpha \left| \text{pass} \right\rangle \left| \text{V} \right\rangle + \beta \left| \text{block} \right\rangle \left| \text{H} \right\rangle).
\end{split}
\end{equation}

Rewriting using the equivalent binary states we get,

\begin{equation}
\begin{split}
&(\alpha \left| \text{0} \right\rangle + \beta \left| \text{1} \right\rangle) \otimes (\lambda \left| \text{0} \right\rangle + \mu \left| \text{1} \right\rangle) \to
\\ &\lambda (\alpha \left| \text{0} \right\rangle \left| \text{0} \right\rangle + \beta \left| \text{1} \right\rangle \left| \text{1} \right\rangle) + \\ &\mu (\alpha \left| \text{0} \right\rangle \left| \text{1} \right\rangle + \beta \left| \text{1} \right\rangle \left| \text{0} \right\rangle).
\end{split}
\end{equation}

We have thus shown in principle how to achieve a fully counterfactual quantum CNOT gate using our dual CQZE setup.

\section{Results and discussion}

\begin{figure}
\centering
\includegraphics[width=0.5\textwidth]{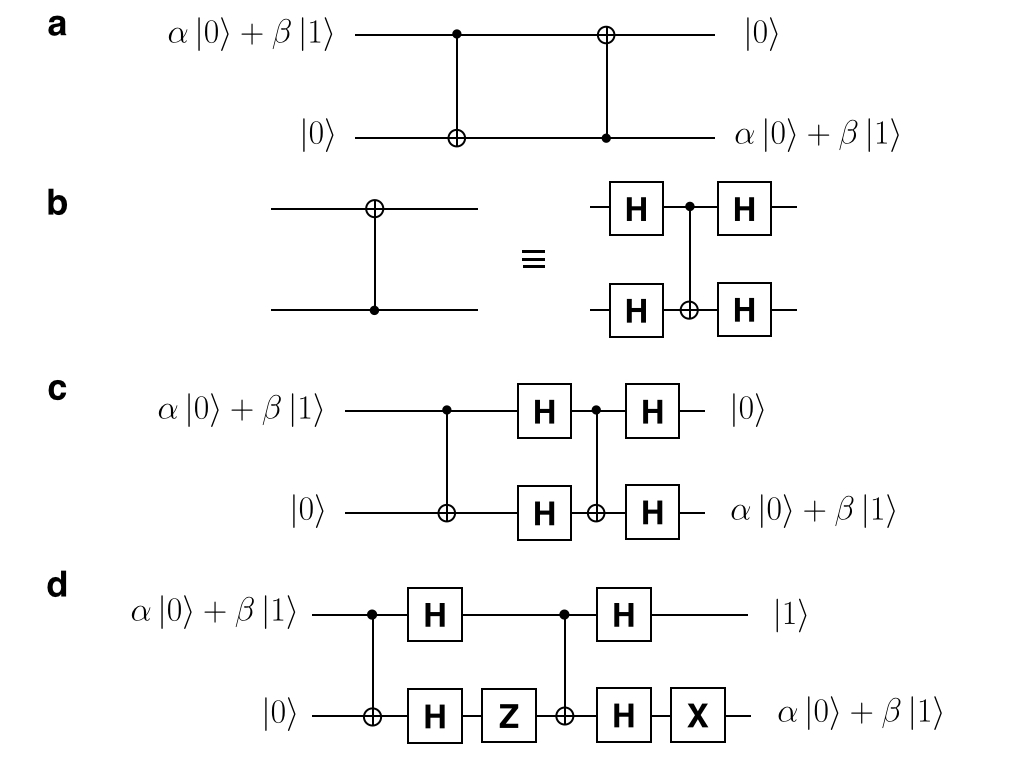}
\caption{\label{fig: DoubleCNOT}{\bf Circuit for quantum state transfer.} {\bf a}, By means of two CNOT gates a qubit $\alpha \left| \text{0} \right\rangle + \beta \left| \text{1} \right\rangle$ can be transferred from one side (top) to the other (bottom). Our counterfactual CNOT gate has Bob's as the control qubit, which is a problem since in this circuit the control qubits of the two CNOT gates are on opposite sides. {\bf b}, By means of four Hadamard gates the control and target qubits of any CNOT gate can be interchanged. {\bf c}, Applying {\bf b} to {\bf a}, the control qubits of both CNOT gates are now on the same side, Bob's. This circuit forms the basis of our protocol for counterfactually transporting an unknown qubit. {\bf d}, In case of a Z-gate before the right hand side CNOT, we need an X-gate at the end in order for the state $\alpha \left| \text{0} \right\rangle + \beta \left| \text{1} \right\rangle$ to be correctly transferred. This circuit is relevant to one of the two possible paths for Alice's exiting photon from the Dual CQZE setup of FIG. \ref{fig: Dual-CQZE}b.}
\end{figure}

We now show how to counterfactually transport an unknown qubit. Using two CNOT gates, the circuit of FIG. \ref{fig: DoubleCNOT}a swaps the input states ($\alpha \left| \text{0} \right\rangle + \beta \left| \text{1} \right\rangle$) and $\left| \text{0} \right\rangle$, effectively transferring $\alpha \left| \text{0} \right\rangle + \beta \left| \text{1} \right\rangle$ from one side (top) to the other (bottom) \cite{Mermin 01}. 

Can we use our counterfactual CNOT gate in this circuit---with Bob's quantum object as the control qubit---to counterfactually transport an unknown state from Bob to Alice? The problem with the circuit is that the control qubits of the two CNOT gates are on opposite sides. But there is a way around it. By means of four Hadamard gates, the circuit of FIG. \ref{fig: DoubleCNOT}b interchanges the control and target qubits of a CNOT gate \cite{Mermin 04}. Applying this to the circuit of FIG. \ref{fig: DoubleCNOT}a we get the circuit of FIG. \ref{fig: DoubleCNOT}c which forms the basis of our protocol,

\begin{equation}
\label{transport circuit}
\begin{split}
&(\alpha \left| \text{0} \right\rangle + \beta \left| \text{1} \right\rangle) \otimes \left| \text{0} \right\rangle \stackrel{\text{CNOT}} {\longrightarrow}
\\ &\alpha \left| \text{00} \right\rangle + \beta \left| \text{11} \right\rangle \stackrel{H^{\otimes 2}} {\longrightarrow}
\\ &\alpha \left| ++ \right\rangle + \beta \left| -- \right\rangle \stackrel{\text{CNOT}} {\longrightarrow}
\\ &\alpha \left| ++ \right\rangle + \beta \left| +- \right\rangle \stackrel{H^{\otimes 2}} {\longrightarrow}
\\ &\left| \text{0} \right\rangle \otimes (\alpha \left| \text{0} \right\rangle + \beta \left| \text{1} \right\rangle).
\end{split}
\end{equation}

What about the case of Alice's exiting photon ending up on path $D_0$, which has $50\%$ probability? From Eq. \ref{Exiting BS}, the combined state of Bob's object and Alice's photon in this path is $\lambda (\alpha \left| \text{pass} \right\rangle \left| \text{H} \right\rangle + \beta \left| \text{block} \right\rangle \left| \text{V} \right\rangle) -
\mu (\alpha \left| \text{pass} \right\rangle \left| \text{V} \right\rangle + \beta \left| \text{block} \right\rangle \left| \text{H} \right\rangle)$, which in binary is $\lambda (\alpha \left| \text{0} \right\rangle \left| \text{0} \right\rangle + \beta \left| \text{1} \right\rangle \left| \text{1} \right\rangle) - \mu (\alpha \left| \text{0} \right\rangle \left| \text{1} \right\rangle + \beta \left| \text{1} \right\rangle \left| \text{0} \right\rangle)$. This is equivalent to the output of a CNOT gate but with Alice applying a $Z$-gate to her input qubit. Incorporating this in the circuit of FIG. \ref{fig: DoubleCNOT}c we get the circuit of FIG. \ref{fig: DoubleCNOT}d. It turns out that an $X$-gate is needed at the end for the state $\alpha \left| \text{0} \right\rangle + \beta \left| \text{1} \right\rangle$ to be transferred from one side (top) to the other (bottom),

\begin{equation}
\begin{split}
&(\alpha \left| \text{0} \right\rangle + \beta \left| \text{1} \right\rangle) \otimes \left| \text{0} \right\rangle \stackrel{\text{CNOT}} {\longrightarrow}
\\ &\alpha \left| \text{00} \right\rangle + \beta \left| \text{11} \right\rangle \stackrel{H^{\otimes 2}} {\longrightarrow}
\\ &\alpha \left| ++ \right\rangle + \beta \left| -- \right\rangle \stackrel{I \otimes Z} {\longrightarrow}
\\ &\alpha \left| +- \right\rangle + \beta \left| -+ \right\rangle \stackrel{\text{CNOT}} {\longrightarrow}
\\ &\alpha \left| -- \right\rangle + \beta \left| -+ \right\rangle \stackrel{H^{\otimes 2}} {\longrightarrow}
\\ &\alpha \left| \text{11} \right\rangle + \beta \left| \text{10} \right\rangle \stackrel{I \otimes X} {\longrightarrow}
\\ &\left| \text{1} \right\rangle \otimes (\alpha \left| \text{0} \right\rangle + \beta \left| \text{1} \right\rangle).
\end{split}
\end{equation}

We have finally arrived at our protocol for counterfactually transporting an unknown qubit.

\emph{Protocol for counterfactual quantum transportation.} Alice starts by sending an $H$-photon into the dual CQZE setup of FIG. \ref{fig: Dual-CQZE}, with Bob's quantum object, his qubit to be counterfactually transported, in a superposition of blocking and not blocking the channel: $\alpha \left| \text{pass} \right\rangle + \beta \left| \text{block} \right\rangle$. Alice then applies a Hadamard transformation to (the polarisation of) her exiting photon, as does Bob to his qubit. Alice sends her photon back into the dual CQZE setup. If her exiting photon is not found in path $D_0$, she knows it is in the other path travelling towards the left. She applies a Hadamard transformation to (the polarisation of) her photon, as does Bob to his qubit. The photon is now in the state $\alpha \left| \text{H} \right\rangle + \beta \left| \text{V} \right\rangle$. If Alice's exiting photon is found instead in path $D_0$, she first applies a Hadamard transformation to (the polarisation of) her photon, as does Bob to his qubit. She then applies an $X$-transformation to her qubit. The photon is now in the state $\alpha \left| \text{H} \right\rangle + \beta \left| \text{V} \right\rangle$. Bob's qubit has been counterfactually transported to Alice. His original qubit ends up in the state $\left| \text{0} \right\rangle$ or $\left| \text{1} \right\rangle$ randomly; in other words destroyed. 

Let us now look at the fidelity of our counterfactual transport as a function of the number of outer and inner cycles $M$ and $N$. We have mentioned that a small number of inner cycles $N$ would lead to output errors for our counterfactual CNOT gate. These output errors only occur for the case of Bob blocking the channel as the number of inner cycles is irrelevant for the case of Bob not blocking the channel. Take the case of Alice sending an $H$-photon into the counterfactual CNOT gate, with Bob blocking. For asymptotically large $N$, with $N \gg M$, and given perfect implementation, the counterfactual CNOT outputs a $V$-photon. However, for finite $N$ we have the recursion relations \cite{Salih},

\begin{equation}
\epsilon[m, N] = \cos (\frac{\pi }{2M})\epsilon[m - 1] - \sin (\frac{\pi }{2M})\cos^N (\frac{\pi }{2N})\eta[m - 1].
\end{equation}  

\begin{equation}
\eta[m, N] = \sin (\frac{\pi }{2M})\epsilon[m - 1] + \cos (\frac{\pi }{2M})\cos^N (\frac{\pi }{2N})\eta[m - 1].
\end{equation}    

where $\eta[M, N]$ is the unnormalised probability amplitude for the photon exiting in the correct state $\left| \text{V} \right\rangle$, and $\epsilon[M, N]$ is the unnormalised probability amplitude for the photon exiting in the incorrect state $\left| \text{H} \right\rangle$. We have the initial conditions $\eta[0, N]=0$, and $\epsilon[0, N]=1$. The case of Alice sending a $V$-photon into the counterfactual CNOT gate is analogous. Using equation \ref{transport circuit}, and bearing in mind that for the component of the superposition where Bob does not block, the probability amplitude of Alice's photon is multiplied by a factor of ${\cos}{\theta}_{M}$ after each outer cycle, we get the following expression for the fidelity of counterfactual qubit transport,

\begin{equation}
%Fidelity=\frac{(1+\eta[M,N])^2}{4}.
Fidelity=\left [ \frac{\alpha^2}{2}{\cos}^M{\theta}_{M}({\cos}^M{\theta}_{M}+\eta [M,N] \pm \epsilon [M,N])+\frac{\beta^2 }{2}\eta [M,N]({\cos}^M{\theta}_{M}+\eta [M,N] \mp \epsilon [M,N]) \right ]^2.
\end{equation} 

\begin{figure}
\centering
\includegraphics[width=0.5\textwidth]{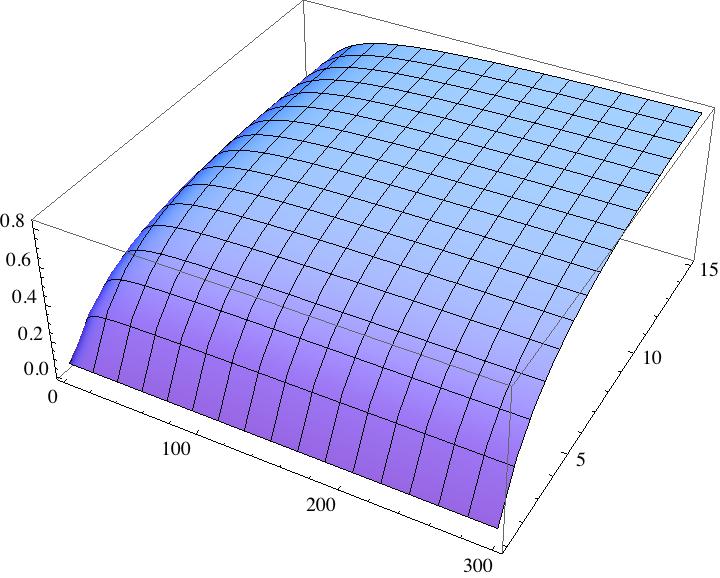}
\caption{\label{fig: Fidelity}{\bf Ideal case fidelity of counterfactual transport.} The fidelity of our protocol for counterfactually transporting an unknown qubit plotted against the number of outer and inner cycles, $M$ and $N$, for $M$ up to 15 and $N$ up to 300. Bob's control qubit is assumed to be an equal superposition here.. Fidelity approaches unity for $N \gg M \gg 1$. Implementation imperfections are ignored.}
\end{figure}

For example, for $M=25$ and $N=320$, with $\alpha = \beta = \frac{1}{\sqrt{2}}$, we already have a fidelity above $86\%$. FIG. \ref{fig: Fidelity} plots fidelity, with $\alpha = \beta = \frac{1}{\sqrt{2}}$, for $M$ up to 15 and $N$ up to 300.

We have so far not said anything about how Bob may practically implement his qubit. Tremendous recent advances mean that there are several candidate technologies. Perhaps most promising for our purpose here are trapped-ion techniques \cite{Hosten, Monroe, Winefield} whereby a carefully shielded and controlled ion can be placed in a quantum superposition of two spatially separated states---one of which in our case blocks the channel. Trapped ions offer relatively long decay times, needed for a large-number-of-cycles implementation of the protocol. Moreover, the Hadamard transformation, key to this protocol, can be directly applied by means of suitable laser pulses. Note that while in our protocol a smaller number of inner and outer cycles, as explained above, leads to higher photon loss and lower fidelity, it opens the door for practical demonstration by post-selecting Alice's photon not lost. Such an implementation also helps with decoherence of Bob's quantum object by shortening the time required for qubit transport. It also helps with errors caused by imperfect optical elements and by noise in the channel---as a larger number of cycles would amplify such errors. Shorter qubit transport time could pose a different challenge however. Switchable optical elements such as switchable mirrors and polarisation rotators, implemented in the single-photon regime, need to be turned on or off quickly enough to achieve the precise timing described in the protocol. A balance needs to be struck. The question of whether our protocol is implementable using current technology is an interesting one whose answer has to wait.

We have proposed a protocol for the counterfactual, disembodied transport of an unknown qubit---much like in quantum teleportation except that Alice and Bob do not require previously-shared entanglement nor a classical channel. No physical particles travel between them either. Here, Bob's qubit is gradually ``beamed up'' to Alice. In the ideal asymptotic limit, efficiency and fidelity approach unity as the probability amplitude of the photon being in the channel approaches zero. This brings into sharp focus both the promise and mystery of quantum information.

\section*{Disclosure/Conflict-of-Interest Statement}

The author declares that the research was conducted in the absence of any commercial or financial relationships that could be construed as a potential conflict of interest.

\section*{Acknowledgments}
I thank Sam Braunstein, M. Suhail Zubairy, M. Al-Amri, Zheng-Hong Li, Gilles Puetz and Nicolas Gisin for useful comments received in 2013 on an initial draft. This paper was first posted on the arXiv on April 8, 2014, arXiv:1404.2200. Qubet Research is a start-up in quantum information.

\clearpage

%\section*{Figures}

%%% Use this if adding the figures directly in the mansucript, if so, please remember to also upload the files when submitting your article
%%% There is no need for adding the file termination, as long as you indicate where the file is saved. In the examples below the files (logo1.jpg and logo2.eps) are in the Frontiers LaTeX folder
%%% If using *.tif files convert them to .jpg or .png

%\begin{figure}[h!]
%\begin{center}
%\includegraphics[width=10cm]{logo1}% This is a *.jpg file
%\end{center}
% \textbf{\refstepcounter{figure}\label{fig:01} Figure \arabic{figure}.}{ Enter the caption for your figure here.  Repeat as  necessary for each of your figures }
%\end{figure}

%\begin{figure}
%\begin{center}
%\includegraphics[width=10cm]{logo2}% This is an *.eps file
%\end{center}
%\textbf{\refstepcounter{figure}\label{fig:02} Figure \arabic{figure}.}{ Enter the caption for your figure here.  Repeat as  necessary for each of your figures }
%\end{figure}

%%% If you don't add the figures in the LaTeX files, please upload them when submitting the article.

%%% Frontiers will add the figures at the end of the provisional pdf automatically %%%

%%% The use of LaTeX coding to draw Diagrams/Figures/Structures should be avoided. They should be external callouts including graphics.

\end{document}